\documentclass{iau} 
\usepackage{graphicx}
\usepackage{natbib}

\newcommand\kmps{\mbox{km\,s$^{-1}$}}

\pubyear{2022}
\volume{370}
\jname{Winds of Stars and Exoplanets}
\editors{A.~A. Vidotto, L.~Fossati \& J.~Vink, eds.}

\title[CW Leonis] 
{The porous envelope and circumstellar wind matter of the closest carbon star, CW Leonis}

\author[Hyosun Kim et al]   
{Hyosun Kim$^1$, Ho-Gyu Lee$^1$, Youichi Ohyama$^2$, Ji Hoon Kim$^3$,
Peter Scicluna$^4$, You-Hua Chu$^2$, Nicolas Mauron$^5$ \and Toshiya Ueta$^6$}

\affiliation{$^1$Korea Astronomy and Space Science Institute, 776,
  Daedeokdae-ro, Yuseong-gu, Daejeon 34055, Republic of Korea \\ [\affilskip]
$^2$Institute of Astronomy and Astrophysics, Academia Sinica,
  11F of Astronomy-Mathematics Building, AS/NTU, No.1, Sec. 4,
  Roosevelt Rd, Taipei 10617, Taiwan, R.O.C. \\ [\affilskip]
$^3$SNU Astronomy Research Center, Department of Physics and Astronomy,
  Seoul National University, 1 Gwanak-ro, Gwanak-gu, Seoul 08826, Korea \\ [\affilskip]
$^4$European Southern Observatory, Alonso de Cordova 3107, Santiago RM,
  Chile \\ [\affilskip]
$^5$Laboratoire Univers et Particules, Universite de Montpellier and CNRS,
  Batiment 13, CC072, Place Bataillon, F-34095 Montpellier,
  France \\ [\affilskip]
$^6$Department of Physics and Astronomy, University of Denver,
  2112 E Wesley Ave., Denver, CO 80208, USA}

\begin{document}

\maketitle

\begin{abstract}
  Recent abrupt changes of CW Leonis may indicate that we are witnessing
  the moment that the central carbon star is evolving off the Asymptotic
  Giant Branch (AGB) and entering into the pre-planetary nebula (PPN)
  phase. The recent appearance of a red compact peak at the predicted
  stellar position is possibly an unveiling event of the star, and the
  radial beams emerging from the stellar position resemble the feature
  of the PPN Egg Nebula. The increase of light curve over two decades is
  also extraordinary, and it is possibly related to the phase transition.
  Decadal-period variations are further found in the residuals of light
  curves, in the relative brightness of radial beams, and in the extended
  halo brightness distribution. Further monitoring of the recent dramatic
  and decadal-scale changes of this most well-known carbon star CW Leonis
  at the tip of AGB is still highly essential, and will help us gain a
  more concrete understanding on the conditions for transition between
  the late stellar evolutionary phases.
\keywords{stars: AGB and post-AGB, (stars:) binaries: general, stars: carbon,
(stars:) circumstellar matter, stars: evolution, stars: individual (CW Leonis),
stars: late-type, stars: mass loss, stars: winds, outflows}
\end{abstract}

\firstsection 
              
\section{Introduction}

Many pre-planetary nebulae (PPN) consist of newly-formed inner bipolar/%
multipolar lobes and outer spirals/rings/arcs that are the fossil records
of stellar wind matter accumulated during the asymptotic giant branch (AGB)
phase. The coexistence of two such morphologically distinct circumstellar
structures is a mystery; however, it is widely believed that binaries play
a key role. The most direct clue to resolving the mystery of the shape
transition along stellar phase evolution may be offered by catching the
moment when an AGB star is evolving off the current phase toward the PPN
phase. Recent dramatic changes of CW Leonis likely indicate that we are
witnessing the moment of transition between these late stellar evolutionary
phases.

\section{Previous Views on CW Leonis}

CW Leonis is the closest \citep[distance of about 123 pc;][]{gro12} and the
most well-studied carbon-rich AGB star (or carbon star). Multi-wavelength
observations suggest that CW Leonis is likely a binary system \citep[e.g.,]
[]{jef14,dec15}. Its non-concentric ring-like pattern over 200 arcsec is
remarkable \citep{mau00}, which can be modeled by a spiral-shell structure
introduced by an eccentric orbit binary at the center \citep[e.g.,][]{cer15}.
However, neither the carbon star nor the companion has been identified
because of obscuration by the dense circumstellar matter ejected from
this extreme carbon star at the tip of AGB.

Several near-infrared observations were executed with adaptive optics
and speckle interferometry in 1995--2003 \citep{tut00,ost00,wei02,mur05}
achieving high-resolutions ($<0.1$ arcsec) but losing stellar positional
information at the cost of field sizes; several clumps were revealed, but
their relationship with the central star was unclear. The vigorous debate
about which clump corresponds to the carbon star ended in vain through a
monitoring study over 2000--2008 showing that the clumps faded out around
2005 \citep{ste16}. 

In the optical, before 2011, the core region exhibited an extended bipolar
nebula without any distinct point source \citep{han98,ski98,lea06}, from
which this object was thought to have an invisible star residing in a
dusty disk lying perpendicular to the bipolar structure. The bipolar-like
structure, however, disappeared in the latest Hubble Space Telescope images
taken in 2011 and 2016 \citep{kim15,kim21}, suggesting a completely different
view for CW Leonis.

\section{Recent Dramatic Changes and Porous Envelope Scenario}

\begin{figure} 
\begin{center}
 \includegraphics[width=\textwidth]{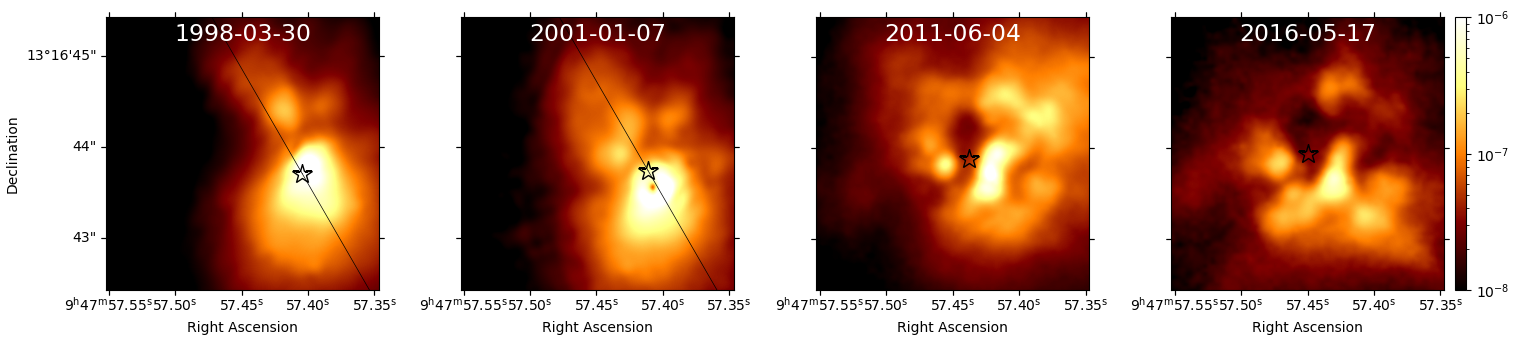}
 \caption{\label{fig:cens}
   Temporal change of brightness in the central 1.5 arcsec region of
   CW Leonis. The star symbol indicates the proper-motion-corrected
   position of the star at each epoch, denoted at the top of each panel.
   The bipolar-like structure (black line) before 2011, which had likely
   given a misleading impression for the origin and evolutionary phase
   of CW Leonis, disappeared in the 2011 and 2016 epochs. From left to
   right, the Hubble Space Telescope images are taken with the F606W
   filter ($\sim0.6\rm\,\mu m$) at the epochs of 1998-03-30 (Prop.~ID:
   6856, PI: J.~Trauger), 2001-01-07 (Prop.~ID: 8601, PI: P.~Seitzer),
   2011-06-04 (Prop.~ID: 12205, PI: T.~Ueta), and 2016-05-17 (Prop.~ID:
   14501, PI: H.~Kim).}
\end{center}
\end{figure}

Surprisingly, the latest optical images of CW Leonis, taken using Hubble
Space Telescope, in 2011 and 2016 revealed dramatic changes in the core
region of the circumstellar envelope from those taken about 10-year earlier
in 1998 and 2001 (Figure\,\ref{fig:cens}). Besides the disappearance of
the long-believed bipolar-like nebula, several important features are
identified and become evidences for a porous envelope of the central star
\citep{kim21}.

\begin{figure} 
\begin{center}
 \includegraphics[width=\textwidth]{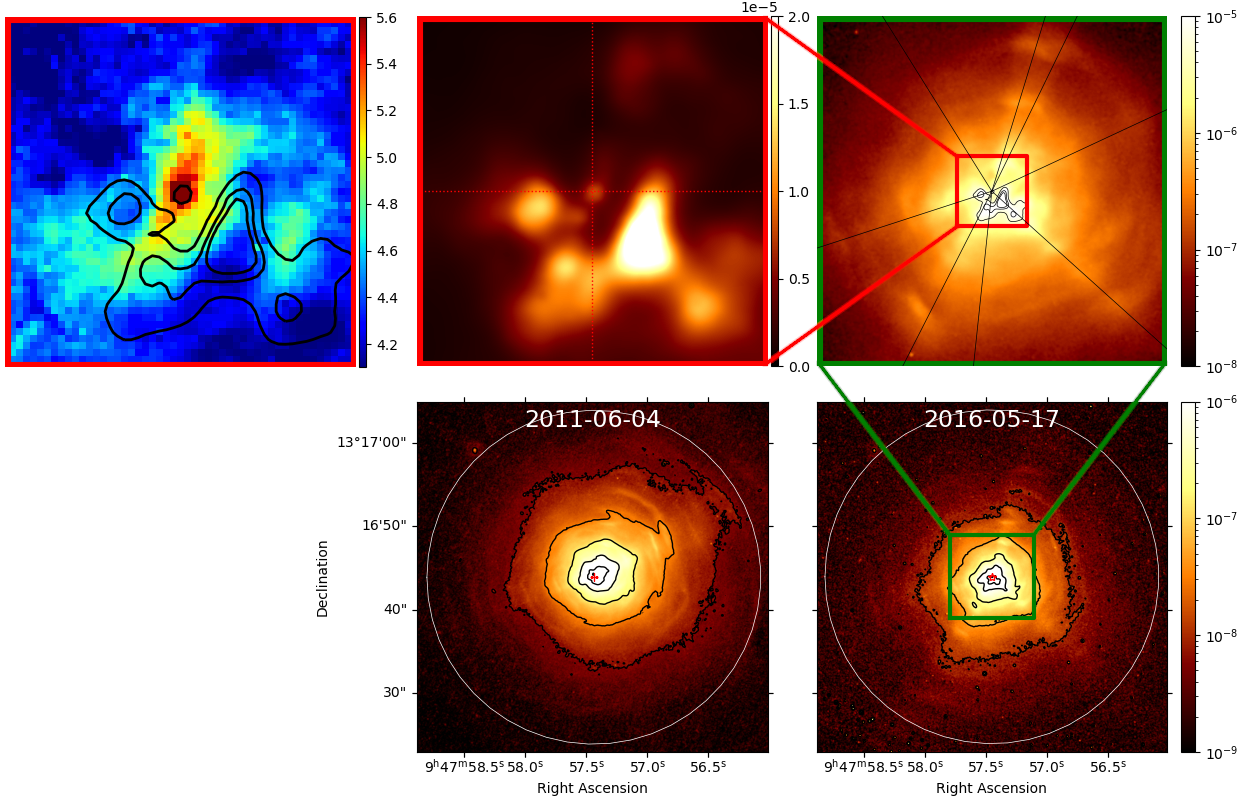}
 \caption{\label{fig:2016}
   The Hubble Space Telescope image of CW Leonis taken with the F814W
   filter ($\sim0.8\rm\,\mu m$) in 2016 (bottom right), compared to the
   2011 epoch image (bottom left). The radial beams and multiple rings
   appearing prominent in the central 5 arcsec region are the intriguing
   features (top right). The color map for central 1 arcsec region shows
   the reddest spot coincides with the stellar position, marked by black
   contours of the F814W brightness (top left) same as the image in the
   top middle panel.
   The color bars range the F814W brightness in logarithmic scales,
   but for the color map in the top left panel being in linear scales
   for the magnitude difference between the F606W and F814W images.}
\end{center}
\end{figure}

In contrast to its absence at the previous epochs, a local brightness peak
appears exactly at the expected stellar position (Figure\,\ref{fig:2016},
top middle) and it is identified as the reddest spot in the color map
(Figure\,\ref{fig:2016}, top left). It is compact; its full width at half
maximum above the adjacent diffuse emission is slightly larger than the
standard point-spread function. This red compact spot at a local peak is
interpreted as the direct starlight, escaping through one of the gaps in
the clumpy envelopes shrouding the star. This radial beam may pulsate
around the line of sight with a small angle and be coincidentally aligned
with the line of sight at the observed epoch in 2016.

The outer part of the observed image exhibits eight straight lines of
brightness that are radially stretched from the central star (Figure%
\,\ref{fig:2016}, top right). In the context of the porous envelope
scenario, these searchlight beams indicate the trajectories of starlight
penetrating the holes in the inner envelopes, along which adjacent dust
particles in the circumstellar envelopes are illuminated. Any other
interpretation for their origin is precluded because of the straightness
of these beams regardless of the considerably fast stellar proper motion.

The extended halo brightness distribution becomes fairly symmetric about
the central star in the 2016 image, compared to the elongated distribution
to the northwest in the 2011 image (Figure\,\ref{fig:2016}, bottom panels).
This change is again beyond the scope of dynamics of matter that would
require a very long time to move the whole halo with an extremely large
extent. It is speculated that one of the radial beams, that is related to
the new emergence of central red compact brightness peak, alters its angle
toward the line of sight in 2016 from a slightly misaligned angle pointing
toward the northwestern direction in 2011, explaining the redistribution
of the halo brightness.

\begin{figure} 
\begin{center}
 \includegraphics[width=\textwidth]{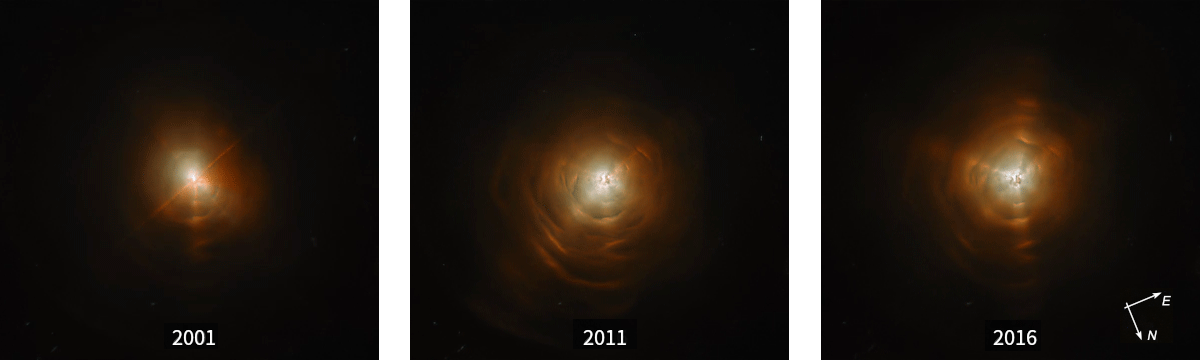}
 \caption{\label{fig:beam}
   Searchlight beams of CW Leonis are fixed in the relative positions
   with respect to the proper-motion-corrected stellar position and their
   relative brightnesses are varying with time. The snapshots are taken
   from an animation in the Research Gallery of http://hubblesite.org.
   Image credit: ESA/Hubble, NASA, Toshiya Ueta (Univ. of Denver), Hyosun
   Kim (KASI).}
\end{center}
\end{figure}

The radial beams appearing in the plane of the sky do not seem to shift
their positions much with time (Figure\,\ref{fig:beam}). Their position
angles with respect to the predicted stellar positions at the individual
epochs are almost fixed. Therefore, it is natural that we assume the
precession angle for the radial beam relevant to the extended brightness
distribution is small. In order to verify this scenario, another epoch
imaging observations with the same setup is anticipated.

In contrast, the relative brightnesses of the radial beams significantly
change and the period seems to be about 10 years (Figure\,\ref{fig:beam}).
The brightest beams are toward north in the 2001 and 2011 epochs (downward
in the figure) while toward south in the 2016 epoch (upward). Indeed, 
besides the stellar pulsation of 640-day period, a decadal variation
has been suggested based on near-infrared and optical photometric data
\citep{dyc91,kim21}. In particular, the increases of $K$-band flux and
the point source contribution in it during 1980--1990 are quite similar
to the event found in 2016. To assess whether the variations are indeed
periodic, more frequent longer-term monitoring observations are desired. 

\section{Anisotropic Wind Expansion induced by an Eccentric Binary}

\begin{figure} 
\begin{center}
 \includegraphics[width=0.5\textwidth]{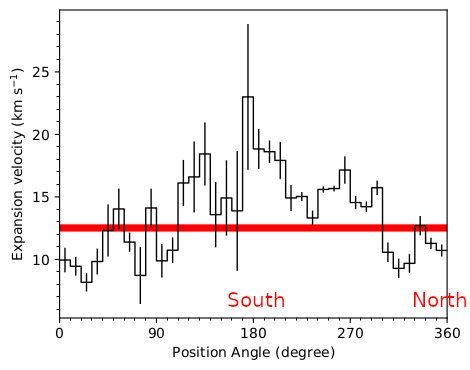}%
 \includegraphics[width=0.5\textwidth]{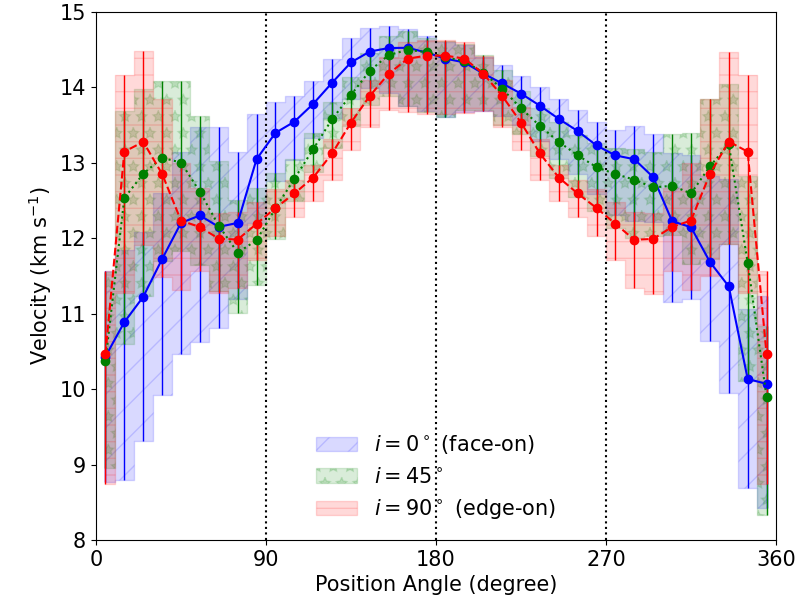}
 \caption{\label{fig:vexp}
   (Left) the expansion velocities in 36 sectors derived from differential
   proper motion of recurrent ring-like pattern of CW Leonis \citep{kim21};
   (right) the correspondance in an eccentric-orbit binary model with the
   pericenter of the mass-losing star to the north (at the position angle
   of 0$^\circ$), viewed at three different inclination angles 0$^\circ$
   (blue, solid line, hatched area), 45$^\circ$ (green, dotted line,
   star-filled area), and 90$^\circ$ (red, dashed line, horizontally
   lined area), respectively \citep{kim22}.}
\end{center}
\end{figure}

The multiple shell pattern wrapping around the star is one of the most
intriguing features of CW Leonis. Our analysis using differential proper
motion of the pattern indicates expansion of shells of ejected material
from the star. The derived speeds of the expanding shells depend on the
direction (Figure\,\ref{fig:vexp}, left). The expansion speeds vary not
only across different position angles within the Hubble Space Telescope
image (about 7\,\kmps\ faster to the south), but their average speed in
the plane of the sky was also about 2\,\kmps\ slower than the wind speed
along the line of sight that is derived from molecular line observations
in radio wavelengths. This variation of measured speeds indicates an
overall nonspherical geometry of the wind matter. We further find that
these observations (Figure\,\ref{fig:vexp}, left panel) are compatible
with a binary model having an eccentric orbit (see the right panel of
Figure\,\ref{fig:vexp}, from \citealp{kim22}).

We regard, however, the velocity measurement was somewhat uncertain. The
2016 image was not as deep as the 2011 image, reducing the number of rings
used in the analysis. Another obstacle was the relatively small expansion
length (the average of positional difference of individual rings between
the 2011 and 2016 epochs) due to the short 5-year interval, which was only
slightly larger than the size of point-spread function. With these reasons,
high-resolution high-sensitivity imaging monitoring is further needed.

\section{Conclusion}

The recent drastic changes in optical images suggest that the previously-seen
bipolar-like structure could not be a concrete structure but could possibly
be parts of searchlight beams with varying relative brightnesses along time.
These radial beams reveal the pathways of starlight illuminating dusty material
after escaping through the gaps in the clumpy envelope enshrouding the star.
The appearance of a distinct brightness peak exactly at the predicted
stellar position and the abnormal shift of large halo distribution both
can be explained by a hypothesized radial beam toward us that is slowly
precessing with a small angle and aligned with the line of sight at the
latest observation epoch.

Although the complexity of CW Leonis has been well known since its discovery,
the three dimensional morphology of its central core and evolving beams is
uniquely revealed with recent Hubble Space Telescope optical monitoring of
the core images. It also allows to trace the expansion velocity of shells
and clumps as seen in the evolving light of the central star. Furthermore,
on-going efforts on three dimensional hydrodynamic models fitting the data
likely suggests a small inclination of the orbit (close to face-on) with the
pericenter of the mass-losing star at North. Further systematic monitoring
of this canonical high mass-loss carbon star is mandatory to strengthen our
interpretation coupled with our modelling. Hubble Space Telescope allows us
to be very near of establishing a robust understanding of the mass loss of
this strategic but mysterious AGB (or soon post-AGB) star.

\acknowledgments
HK acknowledges support by the National Research Foundation of Korea (NRF)
grant (No.\ 2021R1A2C1008928) and Korea Astronomy and Space Science Institute
(KASI) grant (Project No. 2022-1-840-05), both funded by the Korea Government
(MSIT).

\def\apj{{ApJ}}    
\def\nat{{Nature}}    
\def\jgr{{JGR}}    
\def\apjl{{ApJ Letters}}    
\def\aap{{A\&A}}   
\def\mnras{{MNRAS}}
\def\aj{{AJ}}
\def\apjs{{ApJS}}
\let\mnrasl=\mnras


\end{document}